\documentstyle[11pt]{article}

\begin{document}

\title{{\bf Secondly Quantized Multi-Configurational Approach for Atomic
Databases}}
\author{Gediminas Gaigalas and Zenonas Rudzikas \\
        {\em Institute of Theoretical Physics and Astronomy,} \\ 
        {\em A. Go\v{s}tauto 12, Vilnius 2600, Lithuania}}
\date{}
\maketitle


\section{Introduction}

Studies of the structure of atoms and ions (ultracold to relativistic
included) as well as their two-body interactions with photons,
electrons and other particles require accurate methods for the
description of such objects.

In order to obtain accurate values of atomic quantities
it is necessary to account for relativistic and correlation effects.
Relativistic effects may be taken into account as Breit-Pauli
corrections or in a fully relativistic approach. In both cases for
complex atoms and ions, a considerable part of the effort must be
devoted to integrations over spin-angular variables,
occurring in the matrix elements of the operators under consideration.

Many existing codes for integrating
are based on a scheme by Fano \cite{F}.
The integrations over spin-angular variables in this case constitute a
considerable part of the problem, especially when atoms with many open
shells are treated, and the operators are not trivial. In the papers of
Gaigalas {\em et al} \cite{GR,GRF},
the efficient approach for finding
matrix elements of any
one- and two-particle atomic operator between complex configurations is
suggested. It is free of shortcomings of previous approaches.
This approach allows one to generate fairly accurate databases of atomic
parameters (Froese Fischer {\em et al} \cite{FYG,FG}).

 Further development of the approaches by Gaigalas {\em et al} \cite{GR,GRF}
for the spin-spin
and spin-other-orbit relativistic
corrections in the Breit-Pauli approximation is presented in
this poster.

\section{Matrix Elements Between Complex Configurations}

According to the approach of Gaigalas {\it et al} \cite{GRF},
a general expression of the submatrix element
for any two-particle operator between functions with $u$ open shells, can be
written as follows:
\begin{equation}
\label{eq:mg}
\begin{array}[b]{c}
(\psi _u^{bra}\left( LS\right) ||G||\psi _u^{ket}\left( L^{\prime }S^{\prime
}\right) )= \\
=\displaystyle {\sum_{n_il_i,n_jl_j,n_i^{\prime }l_i^{\prime },n_j^{\prime
}l_j^{\prime }}}
(\psi _u^{bra}\left( LS\right) ||\widehat{G}
\left( n_il_i,n_jl_j,n_i^{\prime }l_i^{\prime },n_j^{\prime }
l_j^{\prime } \right)
||\psi _u^{ket}\left( L^{\prime }S^{\prime
}\right) )= \\
=
\displaystyle {\sum_{n_il_i,n_jl_j,n_i^{\prime }l_i^{\prime },n_j^{\prime
}l_j^{\prime }}}\displaystyle {\sum_{\kappa _{12},\sigma _{12},\kappa
_{12}^{\prime },\sigma _{12}^{\prime }}}\sum \left( -1\right) ^\Delta \Theta
^{\prime }\left( n_i\lambda _i,n_j\lambda _j,n_i^{\prime }\lambda _i^{\prime
},n_j^{\prime }\lambda _j^{\prime },\Xi \right) \times \\ \times T\left(
n_i\lambda _i,n_j\lambda _j,n_i^{\prime }\lambda _i^{\prime },n_j^{\prime
}\lambda _j^{\prime },\Lambda ^{bra},\Lambda ^{ket},\Xi ,\Gamma \right)
\times \\
\times R\left( \lambda _i,\lambda _j,\lambda _i^{\prime },\lambda _j^{\prime
},\Lambda ^{bra},\Lambda ^{ket},\Gamma \right),
\end{array}
\end{equation}
where $~\Lambda ^{bra}\equiv \left( L_iS_i,L_jS_j,L_i^{\prime
}S_i^{\prime },L_j^{\prime }S_j^{\prime }\right) ^{bra}~$ is the array of bra
function shells' terms, and similarly for $\Lambda ^{ket}$ and 
$\lambda \equiv ls$.

Thus, to calculate the spin-angular part of a submatrix element of
this type, one has to obtain:

\begin{enumerate}

\item The recoupling matrix $R\left( \lambda
_i,\lambda _j,\lambda _i^{\prime },\lambda _j^{\prime },\Lambda
^{bra},\Lambda ^{ket},\Gamma \right) $, which has an analytical expression in
term of just 6j- and 9j-coefficients.

\item Submatrix elements
$T\left( n_i\lambda _i,n_j\lambda _j,n_i^{\prime }\lambda _i^{\prime
},n_j^{\prime }\lambda _j^{\prime },\Lambda ^{bra},\Lambda ^{ket},\Xi
,\Gamma \right) $,
for tensorial products of creation/annihilation operators that act upon a
particular electron shell. So, all the advantages of tensorial algebra and
quasispin formalism (Rudzikas \cite{Rudz}) may be efficiently exploited in the
process of their calculation.

\item Phase factor $\Delta $,

\item $\Theta ^{\prime }\left(
n_i\lambda _i,n_j\lambda _j,n_i^{\prime }\lambda _i^{\prime },n_j^{\prime
}\lambda _j^{\prime },\Xi \right) $, which is proportional to the two-electron
submatrix element of operator $\widehat{G}$.

\end{enumerate}

Further development of this approach for the spin-spin and spin-other-orbit
relativistic
corrections in the Breit-Pauli approximation is presented in the
following section.

\section{The Spin-Spin and Spin-Other-Orbit Operators}

The {\em spin-spin} operator $H^{ss}$ itself contains tensorial structure of two
different types, summed over $k$:

\begin{eqnarray}
\label{eq:ss-a}
H^{ss} \equiv \displaystyle {\sum_{k}}
\left[ H_{ss}^{\left( k+1k-12,112\right) } + H_{ss}^{\left(
k-1k+12,112\right) } \right].
\end{eqnarray}

Their submatrix elements are:

\begin{eqnarray}
\label{eq:ss-b}
\left( n_i\lambda _in_j\lambda _j\left\| H_{ss}^{\left( k+1k-12,112\right)
}\right\| n_{i^{\prime }}\lambda _{i^{\prime }}n_{j^{\prime }}\lambda
_{j^{\prime }}\right) =\frac 3{
\sqrt{5}}\sqrt{\left( 2k+3\right) ^{\left( 5\right) }}\times  
\nonumber \\ \times
\left( l_i\left\| C^{\left( k+1\right) }\right\| l_{i^{\prime }}\right)
\left( l_j\left\| C^{\left( k-1\right) }\right\| l_{j^{\prime }}\right)
N^{k-1}\left( n_il_in_jl_j,n_{i^{\prime }}l_{i^{\prime }}n_{j^{\prime
}}l_{j^{\prime }}\right),
\end{eqnarray}

\begin{eqnarray}
\label{eq:ss-c}
\left( n_i\lambda _in_j\lambda _j\left\| H_{ss}^{\left( k-1k+12,112\right)
}\right\| n_{i^{\prime }}\lambda _{i^{\prime }}n_{j^{\prime }}\lambda
_{j^{\prime }}\right) =\frac 3{
\sqrt{5}}\sqrt{\left( 2k+3\right) ^{\left( 5\right) }}\times  
\nonumber \\ \times
\left( l_i\left\| C^{\left( k-1\right) }\right\| l_{i^{\prime }}\right)
\left( l_j\left\| C^{\left( k+1\right) }\right\| l_{j^{\prime }}\right)
N^{k-1}\left( n_jl_jn_il_i,n_{j^{\prime }}l_{j^{\prime }}n_{i^{\prime
}}l_{i^{\prime }}\right),
\end{eqnarray}
where we use a shorthand notation
$\left( 2k+3\right) ^{\left( 5\right) } \equiv \left( 2k+3\right)
\left( 2k+2\right)\left( 2k+1\right)\left( 2k\right)\left( 2k-1\right)$
and radial integral in (3), (4) is defined as in Glass and Hibbert \cite{GH}:

\begin{eqnarray}
\label{eq:m-j}
N^k\left( n_il_in_jl_j,n_{i^{\prime }}l_{i^{\prime }}n_{j^{\prime
}}l_{j^{\prime }}\right) = \nonumber \\
=\frac{\alpha ^2}4\int_0^\infty \int_0^\infty P_i\left( r_1\right) P_j\left(
r_2\right) \frac{r_2^k}{r_1^{k+3}}\epsilon (r_1-r_2)P_{i^{\prime }}\left(
r_1\right) P_{j^{\prime }}\left( r_2\right) dr_1dr_{2,},
\end{eqnarray}

where $\epsilon (x)$ is a Heaviside step-function,

\begin{equation}
\label{eq:h-a}\epsilon (x)=\left\{
\begin{array}{ll}
1 ; & \mbox{ for } x>0, \\ 0 ; & \mbox{ for } x\leq 0.
\end{array}
\right.
\end{equation}

The {\em spin-other-orbit} operator $H^{soo}$ itself contains tensorial structure 
of six different types, summed over $k$:

\begin{equation}
\label{eq:soo-a}
\begin{array}[b]{c}
H^{soo} \equiv \displaystyle {\sum_{k}}
\left[ H_{soo}^{\left( k-1k1,101\right) } + H_{soo}^{\left( k-1k1,011\right) }
+ H_{soo}^{\left( kk1,101\right) } + H_{soo}^{\left( kk1,011\right)
} + H_{soo}^{\left( k+1k1,101\right) } +
H_{soo}^{\left( k+1k1,011\right) }
 \right].
\end{array}
\end{equation}

 Their submatrix elements are:

\begin{eqnarray}
\label{eq:b-b}
\left( n_i\lambda _in_j\lambda _j\left\| H_{soo}^{\left( k-1k1,\sigma
_1\sigma _21\right) }\right\| n_{i^{\prime }}\lambda _{i^{\prime
}}n_{j^{\prime }}\lambda _{j^{\prime }}\right) =2\cdot 2^{\sigma _2}\left\{
\left( 2k-1\right) \left( 2k+1\right) \right. \times \nonumber \\
\times \left. \left( l_i+l_{i^{\prime }}-k+1\right) \left(
k-l_i+l_{i^{\prime }}\right) \left( k+l_i-l_{i^{\prime }}\right) \left(
k+l_i+l_{i^{\prime }}+1\right) \right\} ^{1/2} \times \nonumber \\
\times \left( k\right) ^{-1/2}\left( l_i\left\| C^{\left( k\right) }\right\|
l_{i^{\prime }}\right) \left( l_j\left\| C^{\left( k\right) }\right\|
l_{j^{\prime }}\right) N^{k-2}\left( n_jl_jn_il_i,n_{j^{\prime
}}l_{j^{\prime }}n_{i^{\prime }}l_{i^{\prime }}\right),
\end{eqnarray}

\begin{eqnarray}
\label{eq:b-c}
\left( n_i\lambda _in_j\lambda _j\left\| H_{soo}^{\left( kk1,\sigma _1\sigma
_21\right) }\right\| n_{i^{\prime }}\lambda _{i^{\prime }}n_{j^{\prime
}}\lambda _{j^{\prime }}\right) =-2\cdot 2^{\sigma _2}\left( 2k+1\right)
^{1/2}\left( l_i\left\| C^{\left( k\right) }\right\| l_{i^{\prime }}\right)
\times \nonumber \\
\times \left( l_j\left\| C^{\left( k\right) }\right\| l_{j^{\prime }}\right)
\left\{ \left( k\left( k+1\right) \right) ^{-1/2}\left( l_i\left(
l_i+1\right) -k\left( k+1\right) -l_{i^{\prime }}\left( l_{i^{\prime
}}+1\right) \right) \right. \times\nonumber \\
\times \left\{ \left( k+1\right) N^{k-2}\left( n_jl_jn_il_i,n_{j^{\prime
}}l_{j^{\prime }}n_{i^{\prime }}l_{i^{\prime }}\right) -kN^k\left(
n_il_in_jl_j,n_{i^{\prime }}l_{i^{\prime }}n_{j^{\prime }}l_{j^{\prime
}}\right) \right\} - \nonumber \\
\left. -2\left( k\left( k+1\right) \right) ^{1/2}V^{k-1}\left(
n_il_in_jl_j,n_{i^{\prime }}l_{i^{\prime }}n_{j^{\prime }}l_{j^{\prime
}}\right) \right\},
\end{eqnarray}

\begin{eqnarray}
\label{eq:b-d}
\left( n_i\lambda _in_j\lambda _j\left\| H_{soo}^{\left( k+1k1,\sigma
_1\sigma _21\right) }\right\| n_{i^{\prime }}\lambda _{i^{\prime
}}n_{j^{\prime }}\lambda _{j^{\prime }}\right) =2\cdot 2^{\sigma _2}\left\{
\left( 2k+1\right) \left( 2k+3\right) \right. \times \nonumber \\
\times \left. \left( l_i+l_{i^{\prime }}-k\right) \left( k-l_i+l_{i^{\prime
}}+1\right) \left( k+l_i-l_{i^{\prime }}+1\right) \left( k+l_i+l_{i^{\prime
}}+2\right) \right\} ^{1/2} \times \nonumber \\
\times \left( k+1\right) ^{-1/2}\left( l_i\left\| C^{\left( k\right)
}\right\| l_{i^{\prime }}\right) \left( l_j\left\| C^{\left( k\right)
}\right\| l_{j^{\prime }}\right) N^k\left( n_il_in_jl_j,n_{i^{\prime
}}l_{i^{\prime }}n_{j^{\prime }}l_{j^{\prime }}\right) .
\end{eqnarray}

The radial integrals  in (\ref{eq:b-b})-(\ref{eq:b-d})
are (see Glass and Hibbert \cite{GH}):

\begin{eqnarray}
\label{eq:m-k}
V^k\left( n_il_in_jl_j,n_{i^{\prime }}l_{i^{\prime }}n_{j^{\prime
}}l_{j^{\prime }}\right)  = \nonumber \\
=\frac{\alpha ^2}4\int_0^\infty \int_0^\infty P_i\left( r_1\right) P_j\left(
r_2\right) \frac{r_{<}^{k-1}}{r_{>}^{k+2}}r_2\frac \partial {\partial
r_1}P_{i^{\prime }}\left( r_1\right) P_{j^{\prime }}\left( r_2\right)
dr_1dr_2.
\end{eqnarray}

Now we have all we need (the operators for tensorial structure and their
submatrix elements) for obtaining the value of a matrix element of these
operators for any number of open shells in bra and ket functions. This lets us
exploit all advantages of the approach by Gaigalas {\em et al} \cite{GRF}.

The spin-spin and spin-other-orbit operators itself generally contain tensorial
structure of several different types. Therefore the expression (\ref{eq:mg})
must be used separately for each possible tensorial structure for performing
spin-angular integrations according to \cite{GRF}. Each type of tensorial
structure is associated with a different type of recoupling matrix
$R\left( \lambda
_i,\lambda _j,\lambda _i^{\prime },\lambda _j^{\prime },\Lambda
^{bra},\Lambda ^{ket},\Gamma \right) $ and with different matrix elements of
standard tensorial quantities
$T\left( n_i\lambda _i,n_j\lambda _j,n_i^{\prime }\lambda _i^{\prime
},n_j^{\prime }\lambda _j^{\prime },\Lambda ^{bra},\Lambda ^{ket},\Xi
,\Gamma \right) $.

\section{Conclusions}

The tensorial forms of the general secondly quantized spin-spin interaction
operator (\ref{eq:ss-a}) and spin-other-orbit
interaction operator (\ref{eq:soo-a}) and its submatrix elements
( for spin-spin interaction expressions (\ref{eq:ss-b}), (\ref{eq:ss-c}) and for
spin-other-orbit expressions (\ref{eq:b-b}), (\ref{eq:b-c}) and (\ref{eq:b-d}))
are presented. In calculating its matrix elements between functions with $u$
open shells this allows to exploit all the advantages of method by  Gaigalas
{\em et al} \cite{GRF}:

\begin{enumerate}

\item to obtain both diagonal and off-diagonalelements with respect to the
configuration matrix elements in a unified approach,

\item to use in practical
applications the tables of submatrix elements of standard quantities, which
here are both the coordinate representation and the occupation number
representation tensorial operators,

\item to apply the quasispin formalism the
for occupation numbers parts and make use of it,

\item to make use of having
recoupling matrices simpler than in other known approaches.

\end{enumerate}



\end{document}